\documentclass[english,a4,twocolumn]{revtex4-1}
\usepackage[T1]{fontenc}
\usepackage[latin9]{inputenc}
\usepackage[active]{srcltx}
\usepackage{amsmath}
\usepackage{amssymb}
\usepackage{graphicx}
\usepackage{esint}

\makeatletter
\@ifundefined{textcolor}{}
{%
 \definecolor{BLACK}{gray}{0}
 \definecolor{WHITE}{gray}{1}
 \definecolor{RED}{rgb}{1,0,0}
 \definecolor{GREEN}{rgb}{0,1,0}
 \definecolor{BLUE}{rgb}{0,0,1}
 \definecolor{CYAN}{cmyk}{1,0,0,0}
 \definecolor{MAGENTA}{cmyk}{0,1,0,0}
 \definecolor{YELLOW}{cmyk}{0,0,1,0}
}

\makeatother

\usepackage{babel}
\begin{document}

\title{Exact Stochastic Unraveling of an Optical Coherence Dynamics by Cumulant
Expansion }

\author{Jan Ol\v{s}ina$^{1}$, Tobias Kramer$^{2,3}$, Christoph Kreisbeck$^{4}$
and Tom\'{a}\v{s} Man\v{c}al$^{1}$}

\affiliation{$^{1}$Faculty of Mathematics and Physics, Charles University in
Prague, Ke Karlovu 5, CZ-121 16 Prague 2, Czech Republic}

\affiliation{$^{2}$Institut f\"ur Physik, Humboldt-Universit\"at zu Berlin,
12489 Berlin, Germany}

\affiliation{$^{3}$Department of Physics, Harvard University, Cambridge, Massachusetts
02138, USA}

\affiliation{$^{4}$Department of Chemistry and Chemical Biology, Harvard University,
Cambridge, Massachusetts 02138, USA}
\begin{abstract}
A numerically exact Monte Carlo scheme for calculation of open quantum
system dynamics is proposed and implemented. The method consists of
a Monte-Carlo summation of a perturbation expansion in terms of trajectories
in Liouville phase-space with respect to the coupling between the
excited states of the molecule. The trajectories are weighted by a
complex decoherence factor based on the second-order cumulant expansion
of the environmental evolution. The method can be used with an arbitrary
environment characterized by a general correlation function and arbitrary
coupling strength. It is formally exact for harmonic environments,
and it can be used with arbitrary temperature. Time evolution of an
optically excited Frenkel exciton dimer representing a molecular exciton
interacting with a charge transfer state is calculated by the proposed
method. We calculate the evolution of the optical coherence elements
of the density matrix and linear absorption spectrum, and compare
them with the predictions of standard simulation methods. 
\end{abstract}
\maketitle

\section{Introduction}

The theory of absorption spectra of molecular aggregates depends on
a correct evaluation of the time evolution of the investigated system,
either in a formalism of the wave function or the density operator.
In Condon approximation, i.~e.~when the transition dipole moment
of the investigated system can be assumed independent of the nuclear
degrees of freedom (DOF), one can naturally separate the studied system
into an electronic one, which is affected by light and treated explicitly,
and a nuclear bath, which is treated by some effective theory and
does not directly interact with light. Linear absorption spectrum
requires an evaluation of the time-dependent polarization, and consequently
of the optical coherence elements of the electronic density matrix.
For a two level system linearly coupled to a bath of harmonic oscillators,
this problem can be solved exactly for an arbitrary coupling strength
\cite{Doll2008a}. The theory is thus very well established particularly
in the limit of strong system-bath coupling (SBC), when the resonance
coupling between individual transitions in an aggregate can be neglected.
In the weak SBC limit, perturbative treatments in second order can
be applied, leading to various forms of quantum master equations \cite{MayKuehnBook,Zhang1998a,Yang2002b}.
An interesting observation in this context is the convergence of the
second order results and the exact solution for a two level system.
The exact solution is provided by a second order equation \cite{Doll2008a}.

Many practically important cases, especially those related to photosynthetic
antennae, fall in between the two treatable limits. In the weak SBC
limit one can successfully base the theory on the eigenstate basis
of the electronic Hamiltonian, the so-called \emph{excitonic basis}.
In the strong SBC limit, on the other hand, one can rely on the so-called
\emph{local basis}, i.~e.~the basis of states local to the molecules
involved in the aggregate. No such preferred basis is readily available
for an intermediate SBC strength. The intermediate coupling results
in a renormalization of the energy eigenstates of the Hamiltonian
with respect to the excitonic basis, which leads e.~g.~to a presence
of non-zero off-diagonal elements of the stationary long time reduced
density matrix \cite{Mancal2006b,Olsina2010a}. From the point of
view of the reduced density matrix propagation, these off-diagonal
elements are a result of a coupling between equations of motion for
the state populations and the coherences between the basis states,
and/or a coupling between different coherences. The weak coupling
limit is often accompanied by the so-called \emph{secular approximation},
which neglects such coupling. The effects of this coupling are correspondingly
termed \emph{non-secular effects}. In some particular cases, non-secular
effects can be quite dramatic, leading to a noticeable localization
of the eigenstate basis, exhibited e.~g.~by a temperature dependent
absorption band shift from the values corresponding to a delocalized
excitonic basis to those corresponding to a local basis \cite{Renger2004a}.
Similar effects lead to significant changes of fluorescence depolarization
dynamics in photosynthetic antenna as was demonstrated in Ref.~\cite{Kuhn2002a}.
A theory describing correctly the dynamics of the reduced system interacting
with a bath under arbitrary SBC strength would have to be able to
predict a smooth change of the basis in which the long time reduced
density matrix is diagonal from the excitonic bases (weak coupling)
to the local basis (strong coupling). We will denote the basis in
which the reduced density matrix diagonalizes at long times as the
\emph{preferred basis} in analogy with the preferred basis problem
studied in decoherence theory \cite{SchlosshauerBook}. 

Standard approaches to calculation of the dynamics, based on the second
order master equations, often provide unreliable results for non-secular
effects. Outside the secular approximation, one cannot, for example,~guarantee
positivity of the density matrix (see e.~g.~\cite{Olsina2010a}
and the references therein). There is a big variety of different methods
to solve the dynamics of the molecular aggregates apart from the second
order master equation theory. Among stochastic methods we find many
path integral techniques \cite{Stockburger2002,Shi2003a,Xu2005a,Huo2010a},
Monte Carlo methods \cite{Hubbard1959,Beenken2002a,Shao2004a,Zhou2005a,Mo2005a,Lacroix2005a,Shao2006a},
and also the stochastic wavefunction method \cite{Dalibard1992a},
quantum state diffusion method \cite{Aharonov1993a}, multi-configuration
time-dependent Hartree method \cite{Schulze2014}, mixed quantum-classical
Liouville equation \cite{bai2014a} and some other approaches \cite{Mohseni2008a}.
In the formalism of master equations, the Nakajima-Zwanzig identity
\cite{Nakajima1958,Zwanzig1964a} and the Hashitsumae-Shibata-Takahashi
(also known as convolutionless) identity \cite{Hashitsume1977a,Shibata1977a}
are often used as a starting point of higher order perturbation expansion
theories \cite{Schroder2007a}. Some exact results can be obtained
for particular model systems, for example for the problem of a single
molecule with one electronic transition. This problem is analytically
solvable through second cumulant of Magnus expansion \cite{Magnus1954a,Kubo1962,Mukamel1983}
exact for harmonic bath \cite{Mancal2010c}. An exact solution for
a system of multiple coupled transition can be obtained by the Hierarchical
equations of motion (HEOM) \cite{Tanimura1989a,Ishizaki2005a,Tanaka2009}.
Both, path integral methods and HEOM reach the preferred basis in
the long-time evolution \cite{Dijkstra2010}. 

The HEOM method became very popular recently for calculation of excited
state dynamics in photosynthetic systems as it combines feasibility
with accuracy \cite{Ishizaki2009a,Ishizaki2009b,Ishizaki2009c,Hein2012a,Kreisbeck2012a}.
It has been implemented on modern parallel computers \cite{Strumpfer2012}
and graphics-processing units (GPU-HEOM) \cite{Kreisbeck2011a}. One
limitation of the method is that the calculations become more difficult
with decreasing temperature \cite{Ishizaki2009a}.

In this paper, we propose another method which provides a formally
exact solution to the reduced density matrix problem for harmonic
baths. The method is based on a stochastic unraveling of the equation
of motion for the reduced density matrix in the resonance coupling
term. The leading idea is to cover the resonance coupling term in
Hamiltonian by stochastic unraveling rather than doing it with the
SBC as it is usual in ordinary stochastic methods. The evolution of
the system's state is modeled by an ensemble of trajectories in the
space of the projectors on the states in the system's Hilbert space.
This projector space is known in the theory of non-linear spectroscopy
as the \emph{Liouville space} (see Ref.~\cite{MukamelBook}). Each
trajectory from the ensemble can be assigned a sequence of resonance
coupling-free evolution operators that remains after the unraveling.
The resulting expression is related to the high order non-linear response
functions, and it can be evaluated analytically with the knowledge
of the bath correlation function. The properly weighted sum over trajectories
gives an exact result for the system's reduced dynamics. 

The proposed method can offer an advantage over the existing exact
methods in systems with strong system-bath coupling and comparatively
weak resonance coupling since the strong coupling to the bath does
not increase the computational cost. It can be also very well used
in systems with complicated spectral densities, which require more
work in other methods like the HEOM. 

We apply our method to the case of a heterodimer, motivated by the
works on an interaction between charge transfer (CT) states and the
excitons in photosynthetic reaction center \cite{Renger2004a,Mancal2006b}.
In the absorption spectrum of this system, one can observe a large
blue shift of the lowest energy band with increasing temperature \cite{Huber1998a}.
Previous works used either the local basis \cite{Renger2004a} or
the excitonic basis \cite{Mancal2006b} as a starting point of their
theory. It was concluded that the large reorganization energy of the
CT state is the reason for the temperature dependent shift of the
absorption band. It can be shown that within the weak SBC theory,
the necessary condition for the band shift is the difference of the
reorganization energies of the two involved types of states, the excitonic
and the CT states. In second order theories, the non-secular terms,
and correspondingly the shift, vanishes when the reorganization energies
in the dimer are the same \cite{Mancal2008a}. Because in both Refs.
\cite{Renger2004a,Mancal2006b} the description of the shift is provided
by theories which are effectively outside their respective range of
validity, it is important to compare to the shifts predicted by exact
theories. We therefore concentrate on a dimer in which the one dipole
forbidden local state is characterized by a large reorganization energy
and a zero transition dipole moment (playing thus a role of the CT
state), and the second state is optically allowed, characterized by
a moderate reorganization energy (playing a role of an excitonic state).
Due to a resonant interaction between these two excited states, we
observe two peaks in the absorption spectra, which shift as a function
of the system parameters and various approximations discussed. 

The paper is organized as follows: In Section \ref{sec:Model-System}
we introduce the model dimer and its description. Section \ref{sec:AbsorptionSpect}
is devoted to the theory of absorption spectrum and in Section \ref{sec:Stochastic-Unraveling}
we introduce our stochastic unraveling of the equations of motion.
The application of the unraveling is demonstrated and compared with
other methods for calculation of the reduced dynamics in Section \ref{sec:Discussion}.

\section{Model System and Formalism\label{sec:Model-System}}

We illustrate the proposed method on a molecular dimer which represents
the interacting CT state - exciton system. Considering just the linear
absorption, this problem is equivalent to a standard problem of a
molecular heterodimer, and we will therefore formulate it as such.
The special characteristics of the CT state - exciton problem enter
only through a specific set of parameters used for numerical demonstrations.

Each molecule of the dimer is considered to be a two-level system,
either in the ground electronic state $|g_{m}\rangle$ or the excited
electronic state $|e_{m}\rangle$. Index $m$ numbers the molecules.
We introduce the local basis of collective states $|\bar{g}\rangle=|g_{1}\rangle|g_{2}\rangle$,
$|\bar{e}_{1}\rangle=|e_{1}\rangle|g_{2}\rangle$, $|\bar{e}_{2}\rangle=|g_{1}\rangle|e_{2}\rangle$,
and we remove the double-excited state $|\bar{f}\rangle=|e_{1}\rangle|e_{2}\rangle$,
since it is far off-resonant and does not contribute in calculation
of linear absorption spectrum. The total Hamiltonian $H=H_{0}+H_{J}$
is written as sum of Hamiltonian of non-interacting monomers $H_{0}$
and the resonance coupling interaction term $H_{J}$

\begin{align}
H_{0}= & \;\sum_{n=1}^{2}(\epsilon_{n}+\bar{T}+\bar{V}_{n}(\{Q\}))|\bar{e}_{n}\rangle\langle\bar{e}_{n}|\nonumber \\
 & +(\epsilon_{g}+\bar{T}+\bar{V}_{g}(\{Q\}))|\bar{g}\rangle\langle\bar{g}|,\label{eq:Hamiltonian0}\\
H_{J}= & \;(|\bar{e}_{1}\rangle\langle\bar{e}_{2}|+|\bar{e}_{2}\rangle\langle\bar{e}_{1}|)J,\label{eq:HamiltonianJ}
\end{align}
where $\epsilon_{m}$ are energies of collective states $|\bar{e}_{m}\rangle$,
$\epsilon_{g}$ is the collective ground state energy and $J$ is
the resonance coupling between electronic states. We denote $V_{m}^{g}(Q_{m})$,
$V_{m}^{e}(Q_{m})$ the potential electronic surfaces (PES) of the
bath DOF around molecule $m$ in the ground and the excited states,
respectively, and $T_{n}$ represents the kinetic term of nuclear
DOF around molecule $m$. It is more convenient to work with collective
potential and kinetic terms defined as \begin{subequations}
\begin{align}
\bar{V}_{1}(\{Q\}) & =V_{1}^{e}(Q_{1})+V_{2}^{g}(Q_{2}),\\
\bar{V}_{2}(\{Q\}) & =V_{1}^{g}(Q_{1})+V_{2}^{e}(Q_{2}),\\
\bar{V}_{g}(\{Q\}) & =V_{1}^{g}(Q_{1})+V_{2}^{g}(Q_{2}),\\
\bar{T} & =T_{1}+T_{2}.
\end{align}
\end{subequations} The PES $V_{m}^{g}(Q_{m})$ and $V_{m}^{e}(Q_{m})$
depend on the generalized coordinates $Q_{m}$, and we define a collective
coordinate 
\begin{equation}
\{Q\}=\{Q_{1},Q_{2}\}.
\end{equation}
We split the Hamiltonian, Eq.~(\ref{eq:Hamiltonian0}), into the system,
system-bath and bath parts in a standard manner \cite{Olsina2010a}
\begin{align}
H_{0}= & \; H_{S}+H_{B}+H_{S-B},\\
H_{S}= & \;\sum_{n=1}^{2}(\epsilon_{n}+\langle\bar{V}_{n}(\{Q\})-\bar{V}_{g}(\{Q\})\rangle)|\bar{e}_{n}\rangle\langle\bar{e}_{n}|\nonumber \\
 & \;+\epsilon_{g}|\bar{g}\rangle\langle\bar{g}|,\label{eq:HamiltonianS}\\
H_{B}= & \;(\bar{T}+\bar{V}_{g}(Q))\otimes\hat{1},\\
H_{S-B}= & \;\sum_{n=1}^{2}\Delta\bar{V}_{n}|\bar{e}_{n}\rangle\langle\bar{e}_{n}|.\label{eq:HSB-definition}
\end{align}
The angular brackets denote the bath averaging which is defined as
\begin{equation}
\langle\bullet\rangle=\mathrm{Tr}_{B}\{\bullet w_{eq}\},\label{eq:InitialCondition}
\end{equation}
where the trace is performed over the bath DOF, the density matrix
$w_{eq}$ of the bath is assumed to be of the 
\begin{equation}
w_{eq}=\frac{\exp\left(-H_{B}/k_{B}T\right)}{\mathrm{Tr}_{B}\exp\left(-H_{B}/k_{B}T\right)}
\end{equation}
representing the canonical equilibrium, and the symbol $\bullet$
denotes an arbitrary operator. In the definition of $H_{S-B}$, Eq.~(\ref{eq:HSB-definition}),
we used the so-called energy-gap operator defined as 
\begin{equation}
\Delta\bar{V}_{n}=\bar{V}_{n}-\bar{V}_{g}-\langle\bar{V}_{n}-\bar{V}_{g}\rangle.\label{eq:DeltaPhi_definition}
\end{equation}
The dynamics of the open quantum system is described by the reduced
density matrix (RDM) 
\begin{equation}
\rho_{S}(t)=\mathrm{Tr}_{B}W(t),\label{eq:RDM_def}
\end{equation}
where $W(t)$ is the density matrix of the total system.

Before we proceed with the discussion of the bath model, we introduce
the so-called superoperator formalism, which is advantageous for the
description of open quantum systems. We define Liouville superoperator
(the Liouvillian) as 
\begin{equation}
\mathcal{L}_{X}\bullet=\frac{1}{\hbar}\left[H_{X},\bullet\right]_{-},\label{eq:Liouvillian_def}
\end{equation}
where index $X$ can hold values $0$, $J$, $S$, $S-B$ and $B$.
We also define evolution superoperators 
\begin{equation}
\mathcal{U}_{X}(t)\bullet=U_{X}(t)\bullet U_{X}^{\dagger}(t),
\end{equation}
where $U_{X}(t)=\exp(-iH_{X}\, t/\hbar)$ is an ordinary evolution
operator.

The bath is represented by $N_{\mathrm{HO}}$ harmonic oscillators
in our model, further referred to as harmonic oscillator model (HOM).
We pay special attention to three cases: Case $N_{\mathrm{HO}}=0$
(no bath is present) demonstrates the use of proposed method on calculation
of pure quantum state dynamics. Case $N_{\mathrm{HO}}=1$ can serve
as a good test of the method, because time dynamics of complete density
matrix $W(t)$ can be found explicitly, and we can compare between
the explicit calculation and calculation of reduced dynamics by the
proposed method. Case $N_{\mathrm{HO}}=\infty$ is chosen as a typical
example of an open quantum system with irreversible dynamics. For
$N_{\mathrm{HO}}=0$, we simply put 
\begin{equation}
V_{m}^{g}(Q_{m})=V_{m}^{e}(Q_{m})=T_{m}=0.\label{eq:NoBathCondition}
\end{equation}
For $N_{\mathrm{HO}}=1$, we use definition
\begin{align}
V_{m}^{g}(Q_{m}) & =\hbar\omega_{g_{m}}\left(a_{g_{m}}^{\dagger}a_{g_{m}}+\frac{1}{2}\right),\label{eq:Vgm-def}\\
V_{m}^{e}(Q_{m}) & =\hbar\omega_{e_{m}}\left(a_{e_{m}}^{\dagger}a_{e_{m}}+\frac{1}{2}\right)\label{eq:Vem-def}
\end{align}
with the creation and annihilation operators \begin{subequations}\label{eq:Creation_annihilation_def}
\begin{align}
a_{g_{m}} & =\sqrt{\frac{m_{g_{m}}\omega_{g_{m}}}{2\hbar}}\left(x+\frac{i}{m_{g_{m}}\omega_{g_{m}}}\, p\right),\\
a_{g_{m}}^{\dagger} & =\sqrt{\frac{m_{g_{m}}\omega_{g_{m}}}{2\hbar}}\left(x-\frac{i}{m_{g_{m}}\omega_{g_{m}}}\, p\right),\\
a_{e_{m}} & =\sqrt{\frac{m_{e_{m}}\omega_{e_{m}}}{2\hbar}}\left(x+\frac{i}{m_{e_{m}}\omega_{e_{m}}}\, p+d_{m}\right),\\
a_{e_{m}}^{\dagger} & =\sqrt{\frac{m_{e_{m}}\omega_{e_{m}}}{2\hbar}}\left(x-\frac{i}{m_{e_{m}}\omega_{e_{m}}}\, p+d_{m}\right).
\end{align}
\end{subequations} The potential term of the harmonic oscillator
in the excited state is shifted by $d_{m}$ with respect to its potential
in the ground state. In explicit calculation, we solve the Liouville
-- von Neumann equation $\frac{d}{dt}W(t)=i\mathcal{L}W(t)$ to obtain
dynamics of $W(t)$. The RDM, Eq.~(\ref{eq:RDM_def}), is then obtained
by tracing over DOF of the harmonic oscillator. In a calculation by
the proposed Monte-Carlo method, however, we treat the system as open,
i.~e.~the bath is described via the so-called \textit{energy gap
correlation functions} (EGCF) \cite{MukamelBook} 
\begin{equation}
C_{mn}(t)=\langle U(t)\Delta\bar{V}_{m}(\{Q\})U^{\dagger}(t)\Delta\bar{V}_{n}(\{Q\})\rangle,
\end{equation}
and their integrals -- the lineshape functions
\begin{equation}
g_{mn}(t)=\int\limits _{0}^{t}d\tau\int\limits _{0}^{\tau}d\tau'\; C_{mn}(\tau').\label{eq:lineshapef}
\end{equation}
We use the energy gap correlation function \cite{MukamelBook} 
\begin{equation}
C_{mn}(t)=\hbar\lambda\omega\delta_{mn}\left[\coth(\beta\hbar\omega/2)\cos(\omega t)-i\sin(\omega t)\right]\label{eq:EGCF-underdamped}
\end{equation}
for $N_{\mathrm{HO}}=1$. This can be derived from the explicit form
of bath operators, Eqs.~(\ref{eq:Vgm-def}-\ref{eq:Creation_annihilation_def}),
if we assume $m_{g_{m}}=m_{e_{m}}=m$, $\omega_{g_{m}}=\omega_{e_{m}}=\omega$,
and $d_{m}=d$. The parameter 
\begin{equation}
\lambda=\frac{1}{2}d^{2}m\omega^{2}
\end{equation}
 represents the reorganization energy of the system, and $\beta=(k_{B}T)^{-1}$
is the thermodynamic beta with temperature $T$. The EGCF, Eq.~(\ref{eq:EGCF-underdamped}),
provides no relaxation or temperature; $T$ is the temperature of
the initial condition $|\bar{g}\rangle\langle\bar{g}|w_{eq}$ from
which the system gets excited. Note, that in our manuscript, we use
the convention in which reorganization energy has dimension of energy
and the EGCF has dimension of its square. In spectroscopy, it is also
very common to use frequency that corresponds to the reorganization
energy in place of $\lambda$ and EGCF has dimension of frequency
squared. Both conventions differ by the factor $\hbar$ in appropriate
power.

For the case $N_{\mathrm{HO}}=\infty$, we use EGCF of an overdamped
harmonic bath \cite{MukamelBook} 
\begin{align}
C_{mn}(t)= & \hbar\lambda\Lambda\delta_{mn}\left[\cot(\Lambda\hbar\beta/2)-i\right]\exp(-\Lambda t)\nonumber \\
 & +\frac{4\lambda\Lambda\delta_{mn}}{\beta}\sum_{n=1}^{\infty}\frac{\nu_{n}\exp\left(-\nu_{n}t\right)}{\nu_{n}^{2}-\Lambda^{2}}\;.\label{eq:EGCF-overdamped}
\end{align}
 The parameter $\Lambda=1/\tau_{C}$ is given by the characteristic
time of damping of the oscillators $\tau_{C}$. The so-called Matsubara
frequencies are defined as $\nu_{n}=2\pi n/\beta\hbar$.

\section{Theory of Absorption Spectra\label{sec:AbsorptionSpect}}

Linear optical properties of a given system are fully characterized
by the Fourier transform of the linear response function $S^{(1)}(t)$
\cite{MukamelBook}. In particular, the linear absorption coefficient
can be evaluated as 
\begin{equation}
\kappa_{a}(\omega)=\frac{4\pi\omega}{n(\omega)c}\;\mathrm{Re}\int\limits{\rm d}t\, S^{(1)}(t)e^{i\omega t},
\end{equation}
where $n(\omega)$ is frequency-dependent refractive index and $c$
is the speed of light. Let us describe the interaction of the dimer
system with electric field $\mathbf{E}(t)$ by semi-classical Hamiltonian
in dipole approximation
\begin{equation}
H_{SL}(t)=-\boldsymbol{\mu}\cdot\mathbf{E}(t),
\end{equation}
where we introduced the transition dipole moment operator $\boldsymbol{\mu}$.
We rewrite $\mathbf{E}(t)=\boldsymbol{e}E(t)$ using its absolute
value $E(t)$ and the polarization vector $\boldsymbol{e}$. Now,
we can express the response function as 
\begin{equation}
S^{(1)}(t)=\frac{i}{\hbar}\mathrm{Tr}_{B}\left\{ \mu\mathcal{U}(t)[\mu,\rho_{0}]_{-}\right\} ,
\end{equation}
where we introduced the transition dipole moment operator projected
on the polarization vector 
\begin{equation}
\mu=\boldsymbol{\mu}\cdot\boldsymbol{e}=\mu_{eg}|e\rangle\langle g|+\mu_{ge}|g\rangle\langle e|,
\end{equation}
and where $\rho_{0}$ is the initial density matrix of the system.

We can notice that all we need to calculate the absorption spectrum
of the molecular dimer are the coherence elements $\mathcal{U}_{e_{m}g,e_{n}g}(t)$
of the evolution superoperator. The elements $m\neq n$ are often
neglected which corresponds to the secular approximation. In Section
\ref{sec:Discussion}, where we compare our method with standard methods,
we plot directly these elements of evolution superoperator.

\section{Stochastic Unraveling of Coherent Dynamics with Pure Dephasing\label{sec:Stochastic-Unraveling}}

\subsection{Basic Principle}

\label{sub:Introdutcion-BasicPrinciple}We will start with the closed
system, whose evolution superoperator is a solution of the Liouville
equation 
\begin{equation}
\frac{d}{dt}\mathcal{U}(t)=i(\mathcal{L}_{0}+\mathcal{L}_{J})\mathcal{U}(t)\;.
\end{equation}
The reason for separation of $\mathcal{L}$ into $\mathcal{L}_{0}$
and $\mathcal{L}_{J}$ is that for the case $\mathcal{L}_{J}=0$,
we can solve the problem with the bath exactly via the cumulant expansion
\cite{Kubo1962,Mukamel1983}. Provided there is initially no entanglement
between the system and the bath, we can write $\mathcal{U}(t)$ as
a time-ordered exponential using time-dependent perturbation theory
\begin{align}
\mathcal{U}(t) & =\mathcal{U}_{0}(t)\bigg[1-i\int\limits _{0}^{t}d\tau\;\mathcal{L}_{J}^{(int)}(\tau)+\nonumber \\
 & +i^{2}\int\limits _{0}^{t}d\tau\int\limits _{0}^{\tau}d\tau'\;\mathcal{L}_{J}^{(int)}(\tau)\mathcal{L}_{J}^{(int)}(\tau')\dots\bigg]\label{eq:perturbation-series}
\end{align}
The interaction picture is taken with respect to the $\mathcal{U}_{0}(t)$,
i.~e.~$\mathcal{L}_{J}^{(int)}(t)\equiv\mathcal{U}_{0}^{\dagger}(t)\mathcal{L}_{J}\mathcal{U}_{0}(t)$.
The assumption about the non-entangled initial state is quite natural
for systems which have an optical energy gap and reside in the equilibrium
state corresponding to electronic ground state before they are excited
by a short laser pulse.

The proposed stochastic scheme is the following: We generate trajectories,
where system exhibits random jumps between projectors on electronic
states on Liouville space of electronic states. The jumps are generated
in such a way that they reconstruct the action of electronic $J$-coupling,
i.~e.~of $\mathcal{L}_{J}$. Between the jumps, the system evolves
according to $\mathcal{U}_{0}(t)$. We introduce a discretization
of the time axis into intervals $\Delta t$. The model is exact in
the limit $\Delta t\to\infty$. In every time step $\Delta t$, there
is a probability $J_{ij}\,\Delta t/\hbar$ of the jump between states
$|e_{i}\rangle\langle e_{k}|\to|e_{j}\rangle\langle e_{k}|$ and the
same probability of the jump $|e_{k}\rangle\langle e_{i}|\to|e_{k}\rangle\langle e_{j}|$.
In addition to the time evolution according to $\mathcal{U}_{0}(t)$,
the trajectory weighting factor is multiplied by a complex number
$\varphi_{\mathrm{c}}$, which we will call ``coherent factor''
(CF) further on in the text. The coherent factor assures the correct
stochastic unraveling. For each jump between bra-states, the trajectory
gets a factor of $+i$, while for a jump between ket-states, it gets
a factor of $-i.$ Hence
\begin{equation}
\varphi_{\mathrm{c}}=i^{N_{\mathrm{bra}}}(-i)^{N_{\mathrm{ket}}}\;,\label{eq:CohFacDef}
\end{equation}
where $N_{\mathrm{bra}}$ and $N_{\mathrm{ket}}$ are the numbers
of the jumps between bra-states and ket-states in the given trajectory,
respectively. 

If we introduce jump superoperators as \begin{subequations} 
\begin{align}
\mathcal{J}_{\mathrm{bra},i\to j}\bullet & =|e_{j}\rangle\langle e_{i}|\bullet,\\
\mathcal{J}_{\mathrm{ket},i\to j}\bullet & =\bullet|e_{i}\rangle\langle e_{j}|,
\end{align}
\end{subequations} we can describe a trajectory with $N$ jumps in
times $t_{1}$, $\dots$, $t_{N}$ by a sequence of jumps $\mathcal{J}_{\tilde{1}}$,
$\mathcal{J}_{\tilde{2}}$, $\dots$, $\mathcal{J}_{\tilde{N}}$,
where every index $\tilde{k}$ should be replaced by ``details''
of the $k^{\mathrm{th}}$ jump, i.~e.~it should specify if it is
a jump in bra or ket vector, and it should state between which of
the states the jump occurs. The total evolution superoperator can
be then written as
\begin{align}
\mathcal{U}(t)=\frac{1}{N_{\mathrm{tr}}} & \sum_{n=1}^{N_{\mathrm{tr}}}\varphi_{\mathrm{c},n}\;\mathcal{U}_{0}(t-t_{N_{n}})\mathcal{J}_{\tilde{N}_{n},n}\mathcal{U}_{0}(t_{N_{n}}-t_{N_{n}-1})\nonumber \\
 & \times\dots\mathcal{J}_{\tilde{2},n}\mathcal{U}_{0}(t_{2}-t_{1})\mathcal{J}_{\tilde{1},n}\mathcal{U}_{0}(t_{1}-t_{0}).\label{eq:sum-over-trajectories}
\end{align}
The index $n$ numbers the trajectories, and $N_{\mathrm{tr}}$ is
number of trajectories.

To show that Eq.~(\ref{eq:sum-over-trajectories}) gives correct
result for evolution superoperator, we will investigate the individual
terms of the expansion, Eq.~(\ref{eq:perturbation-series}). We can
see that the term ``$\mathcal{U}_{0}(t)1$'' is covered by trajectories
with no jumps, which have the probability 
\begin{equation}
p_{0}=Z^{t/\Delta t},\label{eq:p0-norm}
\end{equation}
where $Z$ is the probability that no jump occurs in time interval
$\Delta t$. If the trajectory starts in a projector $|i_{0}\rangle\langle j_{0}|,$
$Z$ reads as 
\begin{equation}
Z=1-\left(\sum_{n\neq i_{0}}J_{i_{0}n}+\sum_{m\neq j_{0}}J_{mj_{0}}\right)\,\Delta t/\hbar.\label{eq:Z_definition}
\end{equation}
 The term 
\begin{align}
-i\mathcal{U}_{0}(t)\int\limits _{0}^{t}d\tau\; & \mathcal{L}_{J}^{(int)}(\tau)=-i\mathcal{U}_{0}(t)\int\limits _{0}^{t}d\tau\;\mathcal{U}_{0}^{\dagger}(\tau)\mathcal{L}_{J}\mathcal{U}_{0}(\tau)\nonumber \\
 & \approx-i\mathcal{U}_{0}(t)\sum_{n=1}^{t/\Delta t}\mathcal{U}_{0}^{\dagger}(n\Delta t)\mathcal{L}_{J}\mathcal{U}_{0}(n\Delta t)\label{eq:term1-approx}
\end{align}
of Eq.~(\ref{eq:perturbation-series}) is represented by trajectories
with one jump at time $\tau=n\Delta t$, which constitute the individual
terms of the sum in Eq.~(\ref{eq:term1-approx}). A trajectory with
one jump evolves according to the evolution superoperator $\mathcal{U}_{0}(t)=\mathcal{U}_{0}^{i'j}(t)\mathcal{U}_{0}^{i'j\dagger}(\tau)\mathcal{U}_{0}^{ij}(\tau)$,
which corresponds to the time evolution in the projector $|i\rangle\langle j|$
for time $\tau$, the action of $\mathcal{L}_{J}$ which transfers
the projector $|i'\rangle\langle j|$ into the projector $|i\rangle\langle j|$
and a time evolution in this projector for time $t-\tau$. The factor
``$-i$'' (complex unity) is included in the $\varphi_{\mathrm{c}}$.
The probability of such a trajectory with time of the jump $\tau$
is given by
\begin{equation}
p_{i'j,ij}=J_{ii'}\,\Delta t/\hbar\; Z^{t/\Delta t-1}\;.
\end{equation}
It yields the correct ratio 
\begin{equation}
\frac{p_{i'j,ij}}{p_{0}}=\mathcal{L}_{J}^{i'j,ij}Z^{-1}\approx\mathcal{L}_{J}^{i'j,ij}\;.\label{eq:probabilities_proportion}
\end{equation}
The Liouvillian $\mathcal{L}_{J}$ is hence not explicitly present
in the sum over trajectories, but it is included by ratio of trajectories
with particular number of jumps, as
\[
\frac{1}{\hbar}\mathcal{L}_{J}=\sum_{i,j}J_{ij}\left(\mathcal{J}_{\mathrm{bra},i\to j}-\mathcal{J}_{\mathrm{ket},i\to j}\right)\;.
\]
There is also a trajectory with a jump between the projectors $|i\rangle\langle j|$
and $|i\rangle\langle j'|$ for each trajectory with a jump from $|i\rangle\langle j|$
to $|i'\rangle\langle j|$. It comes from the second term of the commutator,
Eq.~(\ref{eq:Liouvillian_def}). The trajectory gets additional minus
sign, and the CF is therefore $i$, see Eq.~(\ref{eq:CohFacDef}).
One easily verifies that trajectories with multiple jumps reconstruct
the higher order terms of the expansion, Eq.~(\ref{eq:perturbation-series}).

\subsection{Bath influence}

For the case of a closed system, the superoperators $\mathcal{U}_{0}(t)$
in Eq.~(\ref{eq:sum-over-trajectories}) are obtained explicitly.
For open systems, we perform a trace over bath DOF and the cumulant
expansion (CE), and we get complex factors in terms of the lineshape
functions, Eq.~(\ref{eq:lineshapef}). The reduced evolution superoperator
can then be written as 
\begin{align}
\mathrm{Tr}_{B}\mathcal{U}(t)=\; & \frac{1}{N_{\mathrm{traj.}}}\sum_{n=1}^{N_{\mathrm{traj.}}}C_{n}\varphi_{\mathrm{c},n}\;\mathcal{U}_{S}(t-t_{N_{n}})\times\nonumber \\
 & \mathcal{J}_{\tilde{N}_{n},n}\mathcal{U}_{S}(t_{N_{n}}-t_{N_{n}-1})\dots\mathcal{J}_{\tilde{1},n}\mathcal{U}_{S}(t_{1}-t_{0})\;.\label{eq:ReducedEvops-with-bath}
\end{align}
The factor 
\begin{equation}
C_{n}=\mathrm{Tr}_{B}\left\{ \mathcal{U}_{B}^{\tilde{N}_{n}}(t_{N_{n}}-t_{N_{n}-1})\dots\mathcal{U}_{B}^{\tilde{1}}(t_{1}-t_{0})w_{eq}\right\} \label{eq:bath-factors}
\end{equation}
can be evaluated analytically using second order cumulant expansion
in a manner similar to the evaluation of non-linear response functions
(see e.~g.~\cite{MukamelBook}).

\subsection{Some Numerical Considerations}

We described the basic principle of the method in the previous section,
and we showed that it is equivalent to the time evolution via the
expansion, Eq.~(\ref{eq:perturbation-series}). The sum over trajectories
gives the evolution superoperator $\mathcal{U}(t)$ in some fixed
time $t$. $\mathcal{U}(t)$ is, however, calculated up to a normalization
constant, which depends linearly on the number of trajectories and
according to Eq.~(\ref{eq:p0-norm}) also on time. We would like
to generate trajectories to a maximum time $t_{\mathrm{max}}$, and
use them to evaluate $\mathcal{U}(t)$ for all times $t<t_{\mathrm{max}}$
in such a way that the trajectories are not generated for each time
independently. One possibility is simply to use scaling of the normalization,
Eq.~(\ref{eq:p0-norm}), for times $t<t_{\mathrm{max}}$ of every
trajectory. For technical reasons, we decided to use a different method.
We included the trajectory in summation only at times $t$ for which
$t_{\mathrm{LJ}}<t<t_{\mathrm{max}}$, where $t_{\mathrm{LJ}}$ is
the time of the last jump in the trajectory, and we ignored the trajectory
in evaluation of the times $t<t_{LJ}$. This also leads to the correct
result, because the ratio of the trajectories that have a jump in
the interval $t_{\mathrm{LJ}}<t<t_{\mathrm{max}}$ to the total number
of trajectories is proportional to the scaling of normalization factor
Eq.~(\ref{eq:p0-norm}) with time

\begin{equation}
\frac{p_{\mathrm{has\, jumps\, in\,}(t_{\mathrm{LJ}},t_{\mathrm{max}})}}{p_{\mathrm{all}}}=Z^{(t_{\mathrm{max}}-t_{\mathrm{LJ}})/\Delta t}.
\end{equation}
Ignoring trajectories at $t<t_{LJ}$ times thus provides a correct
normalization.

\subsection{Connection to the Feynman-Vernon Influence Functional}

The connection can be drawn between the described method and the well-known
Feynman-Vernon influence functional \cite{Feynman1963a,Grabert1988a}.
The path integral for a time evolution of a reduced density matrix
can be written in a time-discretized form on a Hilbert space of a
finite dimension, similarly to Eq.~(5) in \cite{Makri1995a}: 

\begin{align}
\rho(s'',s';t)= & \sum_{s_{0}^{+}}\sum_{s_{1}^{+}}\dots\sum_{s_{N-1}^{+}}\sum_{s_{0}^{-}}\sum_{s_{1}^{-}}\dots\sum_{s_{N-1}^{-}}\nonumber \\
 & \langle s''|e^{-iH_{SJ}\Delta t/\hbar}|s_{N-1}^{+}\rangle\dots\langle s_{1}^{+}|e^{-iH_{SJ}\Delta t/\hbar}|s_{0}^{+}\rangle\nonumber \\
 & \times\langle s_{0}^{+}|\rho_{S}(0)|s_{0}^{-}\rangle\times\nonumber \\
 & \langle s_{0}^{-}|e^{iH_{SJ}\Delta t/\hbar}|s_{1}^{-}\rangle\dots\langle s_{N-1}^{-}|e^{iH_{SJ}\Delta t/\hbar}|s'\rangle\nonumber \\
 & I(s_{0}^{+},\dots,s_{N-1}^{+},s'',s_{0}^{-},\dots,s_{N-1}^{-},s';t)\;,\label{eq:Makri-path-integral-2-1}
\end{align}

\noindent where $H_{SJ}=H_{S}+H_{J}$, $\rho_{S}(0)$ is the initial
system reduced density matrix and $s_{i}^{+}$, $s_{i}^{-}$, $s'$,
$s''$ number states from the system's Hilbert space. The sums run
through the whole Hilbert space of the system. The time is discretized
into $N$ steps of size $\Delta t$. The influence functional has
a form 
\begin{align}
I(s_{0}^{+}, & \dots,s_{N-1}^{+},s'',s_{0}^{-},\dots,s_{N-1}^{-},s';t)\nonumber \\
=\, & \mathrm{Tr}_{B}[e^{-i\langle s''|H_{BSB}|s''\rangle\Delta t/2\hbar}e^{-i\langle s_{N-1}^{+}|H_{BSB}|s_{N-1}^{+}\rangle\Delta t/\hbar}\nonumber \\
 & \times\dots e^{-i\langle s_{0}^{+}|H_{BSB}|s_{0}^{+}\rangle\Delta t/2\hbar}w_{eq.}e^{i\langle s_{0}^{-}|H_{BSB}|s_{0}^{-}\rangle\Delta t/2\hbar}\nonumber \\
 & \times\dots e^{i\langle s_{N-1}^{-}|H_{BSB}|s_{N-1}^{-}\rangle\Delta t/\hbar}e^{i\langle s'|H_{BSB}|s'\rangle\Delta t/2\hbar}]\;,\label{eq:influence-func-1-1}
\end{align}
where $H_{BSB}=H_{B}+H_{S-B}$.

Our method, to which we will refer to as Stochastic Unraveling of
the Resonance Coupling (SURC) in the rest of this work, relies on
an approximation of the expression $e^{-iH_{SJ}\Delta t/\hbar}$ in
the Eq.~(\ref{eq:Makri-path-integral-2-1}) using the Trotter expansion
$e^{-i(H_{S}+H_{J})\Delta t/\hbar}\approx e^{-iH_{S}\Delta t/\hbar}e^{-iH_{J}\Delta t/\hbar}$.
This expression after further approximation yields 
\begin{equation}
e^{-iH_{SJ}\Delta t/\hbar}\approx e^{-iH_{S}\Delta t/\hbar}\sum_{rs}\left(\delta_{rs}-\frac{i}{\hbar}J_{rs}\Delta t\,\right)|r\rangle\langle s|\;.
\end{equation}
If the path integral was to be performed without any importance sampling
at this point, at each time step, we would have trajectories that
jump between the states $|r\rangle$ and $|s\rangle$ and gain factor
$iJ_{rs}\Delta t/\hbar$ and those that stay in the same state $|r\rangle=|s\rangle$
and gain factor ``1''. Trajectories with too many jumps, however,
tend to cancel each other. The SURC can thus be viewed as an importance
sampling in which we prefer the trajectories with less jumps by factor
$J_{rs}\Delta t/\hbar$, and we increase correspondingly the phase
change if the jump occurs. This allows us to trivially perform most
of the summations in Eq.~(\ref{eq:Makri-path-integral-2-1}), because
terms $\langle s_{i}^{-}|e^{iH_{SJ}\Delta t/\hbar}|s_{i+1}^{-}\rangle\approx\langle s_{i}^{-}|e^{iH_{S}\Delta t/\hbar}|s_{i+1}^{-}\rangle\delta_{s_{i}^{-}s_{i+1}^{-}}$
for most of the cases. Luckily, our influence functional has a particularly
simple form: If there is a sequence of consecutive states $|s_{i}\rangle$,
$|s_{i+1}\rangle$,$\dots$, $|s_{n}\rangle$, for which $|s_{i}\rangle=|s_{i+1}\rangle=\dots=|s_{n}\rangle$,
all factors with indices between $(i+1)$ and $(n-1)$ can be expressed
as one factor in terms of the lineshape functions and the states can
be excluded from the expression for the influence functional.

\section{Numerical Demonstrations\label{sec:Discussion} }

In this section, we study the dynamics of optical coherences in a
molecular dimer in order to test the precision and numerical stability
of our method. We will demonstrate how the SURC works for a simple
system with no bath (case $N_{\mathrm{HO}}=0$ of HOM) and for an
exactly solvable problem with simple bath model (case $N_{\mathrm{HO}}=1$
of HOM). We also calculate absorption spectra of a model dimer with
full harmonic bath (case $N_{\mathrm{HO}}=\infty$ of HOM) and compare
the results obtained by our method with those obtained by the time-dependent
Redfield tensor (described in Ref.~\cite{Olsina2010a}), its secularized
version and the HEOM.

\subsection{Hierarchical Equations of Motion}

The HEOM as described in Ref.~\cite{Dijkstra2010} is an exact method
for systems with overdamped bath, Eq.~(\ref{eq:EGCF-overdamped}).
To achieve its convergence at low temperatures becomes increasingly
difficult, since the computational cost increases exponentially with
both the system size and the number $N_{\mathrm{Mat}}$ of included
Matsubara frequencies. We denote the calculations with HEOM method
using a particular number $N_{{\rm Mat}}$ of Matsubara frequencies
by an abbreviation HEOM~$N_{\mathrm{Mat}}$. For practical HEOM calculations
on systems larger than a dimer, it is vital to employ the high-temperature
approximation (HTA) to correlation function of Eq.~(\ref{eq:EGCF-overdamped}).
This yields \cite{Ishizaki2009a} 
\begin{align}
C_{mn}^{\mathrm{HTA}}(t)= & \,\frac{2\lambda}{\beta}\frac{(3\Lambda^{2}-\nu_{1}^{2})e^{-\Lambda t}-2\Lambda\delta(t)}{\Lambda^{2}-\nu_{1}^{2}}\delta_{mn}\nonumber \\
 & -i\hbar\lambda\Lambda\delta_{mn}e^{-\Lambda t}\;,\label{eq:CF-HTA-1}
\end{align}
and it requires $\hbar\beta\Lambda<1$. We denote the calculations
with this method by abbreviation HEOM HTA. Compared to HEOM 0, i.~e.~to
neglecting the Matsubara part of the EGCF completely, the HTA extends
the range of validity of HEOM towards lower temperatures while at
the same time it preserves the crucial exponential time-decay of the
correlation function. This is essential for constructing the reduced
hierarchy in a computationally efficient manner. Arbitrary spectral
densities and lower temperatures are readily implemented in SURC using
the exact EGCF, whereas the treatment by HEOM requires a decomposition
of the spectral density into shifted peaks \cite{Tanimura1994,Kreisbeck2012a}.

\subsection{Stochastic Unraveling of the Resonance Coupling}

\begin{figure}[h]
\begin{centering}
\includegraphics[width=1\columnwidth]{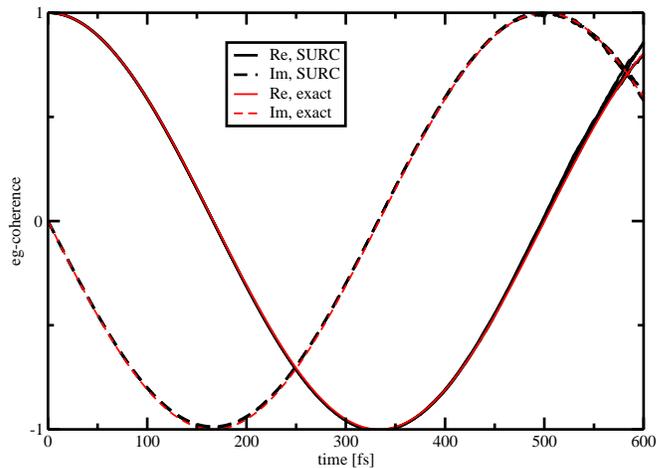}
\par\end{centering}

\caption{``Optical coherence'' element $\mathcal{U}_{e_{1}g,e_{1}g}(t)$
of the evolution superoperator of a molecular homodimer without contact
with a bath. The site and resonance interaction energies are $\epsilon_{1}=\epsilon_{2}=10^{4}\;\mathrm{cm}^{\mbox{-}1}$,
$J=50\;\mathrm{cm}^{\mbox{-}1}$, and the optical frequency $10^{4}\;\mathrm{cm}^{\mbox{-}1}$
is subtracted from the plot. The red lines correspond to the explicit
exact solution of the dynamics, black lines represent the dynamics
calculated by SURC using $10^{8}$ trajectories. Full lines correspond
to the real part and dashed lines to the imaginary part. \label{fig:SchrodingerDynamics}}
\end{figure}
 In order to test the ideas of Section \ref{sec:Stochastic-Unraveling},
we first demonstrate that the method correctly reproduces coherent
quantum dynamics. We set the same excitation energies of both molecules
$\epsilon_{1}=\epsilon_{2}=10^{4}\;\mathrm{cm}^{\mbox{-}1}$ and a
non-zero resonance coupling $J=50\;\mathrm{cm}^{\mbox{-}1}$. 
\begin{figure}[h]
\begin{centering}
\includegraphics[width=1\columnwidth]{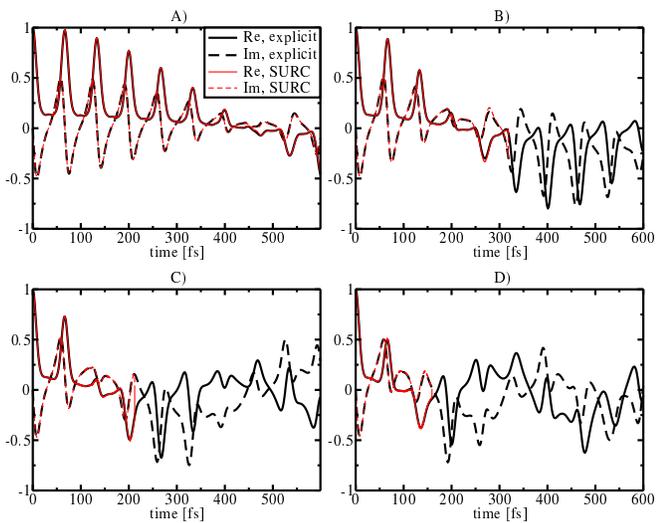}
\par\end{centering}

\caption{``Optical coherence'' element $\mathcal{U}_{e_{1}g,e_{1}g}(t)$
of the evolution superoperator of a molecular homodimer coupled with
one harmonic oscillator per site. The site energies and the reorganization
energies are $\epsilon_{1}=\epsilon_{2}=10^{4}\;\mathrm{cm}^{\mbox{-}1},\lambda_{1}=\lambda_{2}=500\;\mathrm{cm}^{\mbox{-}1}$,
the oscillator frequencies are $\omega_{1}=\omega_{2}=500\;\mathrm{cm}^{\mbox{-}1}$,
and the frequency $10^{4}\;\mathrm{cm}^{\mbox{-}1}$ is subtracted
from the plot. The resonance interaction energy $J$ has values of
$50\;\mathrm{cm}^{\mbox{-}1}$, $100\;\mathrm{cm}^{\mbox{-}1}$, $150\;\mathrm{cm}^{\mbox{-}1}$,
$200\;\mathrm{cm}^{\mbox{-}1}$ in plots A), B), C) and D). The red
lines correspond to the explicit exact solution of the dynamics, black
lines represent dynamics by SURC using $10^{8}$ trajectories. Full
lines correspond to the real part and dashed lines to the imaginary
part. Temperature of the initial condition is $T=100\;\mathrm{K}$.
\label{fig:ExactModelTesting-Dynamics}}
\end{figure}
 The bath is not present, and the dynamics can therefore be solved
exactly. We calculate the element $\mathcal{U}_{e_{1}g,e_{1}g}(t)$
of the evolution superoperator, and we compare the exact dynamics
with the dynamics obtained by SURC. The results are presented in Fig.~\ref{fig:SchrodingerDynamics}
We can see the results of three runs of SURC, each of them performed
with $10^{8}$ trajectories. The time step is chosen so that it smoothly
discretizes the line shape functions, the oscillations given by the
difference between electronic transition energies of the monomers
and the oscillations due to the coupling $J$. For $10^{8}$ trajectories,
numerical noise leads to accuracy breakdown for times longer than
approximately $T_{per.}=2\pi\hbar/J$, and this value is independent
of $dt$ when it is chosen to give a smooth discretization. We chose
to discretize the interval $\langle0;\ T_{per.}\rangle$ into 2048
steps which gives the time step $dt=16.28\;\mathrm{fs}\;\mathrm{cm}^{\mbox{-}1}/J$.
In calculations where the time step would be larger than 0.25~fs,
we chose $dt=0.25\;\mathrm{fs}$ instead. As the Fig.~\ref{fig:SchrodingerDynamics}
demonstrates, SURC dynamics is very close to the exact one before
the dynamics is overwhelmed by noise. 

To show that the SURC method is exact also for an open system with
harmonic bath, we calculate element $\mathcal{U}_{e_{1}g,e_{1}g}(t)$
of the evolution superoperator for a molecular dimer with simple bath
that allows exact solution of the model. The bath is represented by
a single undamped harmonic oscillator (case $N_{\mathrm{HO}}=1$ of
HOM) for each molecule. In the SURC calculation, harmonic oscillators
are treated implicitly by EGCF, Eq.~(\ref{eq:EGCF-underdamped}).
This calculation is compared in Fig.~\ref{fig:ExactModelTesting-Dynamics}
with the exact explicit solution. The calculated system is a homodimer
with parameters $\epsilon_{1}=\epsilon_{2}=10^{4}\;\mathrm{cm}^{\mbox{-}1},\lambda_{1}=\lambda_{2}=500\;\mathrm{cm}^{\mbox{-}1},\omega_{1}=\omega_{2}=500\;\mathrm{cm}^{\mbox{-}1}$
and the temperature of initial condition, Eq.~(\ref{eq:InitialCondition}),
equals to $100\;\mathrm{K}$. We performed four calculations for values
of $J=50\;\mathrm{cm}^{\mbox{-}1}$, $100\;\mathrm{cm}^{\mbox{-}1}$,
$150\;\mathrm{cm}^{\mbox{-}1}$ and $200\;\mathrm{cm}^{\mbox{-}1}$.
We can see that the correspondence of the exact solution with the
SURC solution is very good. The time up to which we can calculate
is again inversely proportional to $J$.

Finally, let us compare the absorption spectra calculated for a bath
model with $N_{HO}=\infty$ by SURC and other frequently used theories.
All the following SURC calculations are performed with the use of
$10^{10}$ trajectories. This prolongs the time before the accuracy
breakdown to approximately $1.5\ T_{per.}$, and it required approximately
2.5 CPU-years of Intel$^{\circledR}$ Xeon$^{\circledR}$ CPU E5620
@ 2.40GHz per calculation. We have chosen the same number of time
steps on the extended interval yielding $dt=23.3\;\mathrm{fs}\;\mathrm{cm}^{\mbox{-}1}/J$.
 We use a dimer model with parameters $\epsilon_{1}=9600\;\mathrm{cm}^{\mbox{-}1},\epsilon_{2}=10000\;\mathrm{cm}^{\mbox{-}1},\lambda_{1}=100\;\mathrm{cm}^{\mbox{-}1},\lambda_{2}=1000\;\mathrm{cm}^{\mbox{-}1},\tau_{C,1}=\tau_{C,2}=100\;\mathrm{fs}$.
The big difference in reorganization energies is typical in situations
where the molecules have very different surroundings (e.~g.~protein
envelope and water) or for systems with both excitonic and CT states,
as discussed in the Introduction. Only the site~1 has non-zero transition
dipole moment, since the site~2 represents a CT state. The length
of the dipole moment is arbitrary since it only changes normalization
of the spectra. 

\begin{figure}
\begin{centering}
\includegraphics[width=1\columnwidth]{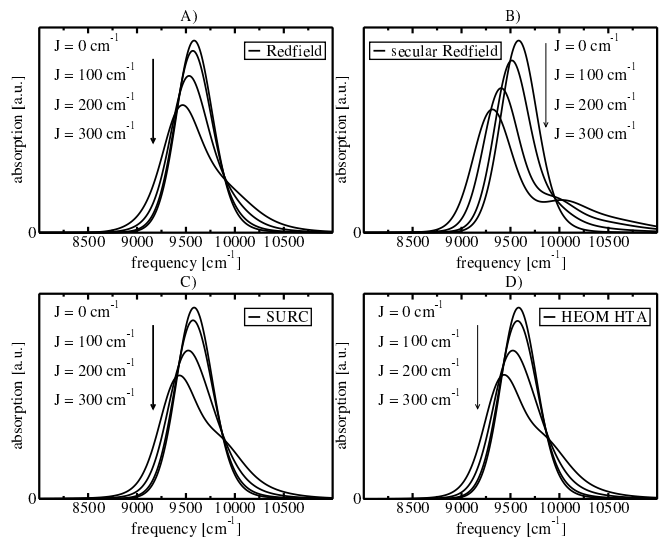}
\par\end{centering}

\caption{Absorption spectra of a molecular homodimer coupled to an overdamped
harmonic bath ($N=\infty$ in HOM) calculated by A) Redfield theory,
B) secular Redfield theory, C) SURC and D) HEOM HTA. The parameters
are $\epsilon_{1}=9600\;\mathrm{cm}^{\mbox{-}1}$ , $\epsilon_{2}=10000\;\mathrm{cm}^{\mbox{-}1}$,
$\lambda_{1}=100\;\mathrm{cm}^{\mbox{-}1}$, $\lambda_{2}=1000\;\mathrm{cm}^{\mbox{-}1}$,
$\tau_{C,1}=\tau_{C,2}=100\;\mathrm{fs}$ and $T=300\;\mathrm{K}$.
Absorption spectra are calculated for $J=0\;\mathrm{cm}^{\mbox{-}1}$
, $J=100\;\mathrm{cm}^{\mbox{-}1}$, $J=200\;\mathrm{cm}^{\mbox{-}1}$
and $J=300\;\mathrm{cm}^{\mbox{-}1}$. \label{fig:Absorption-spectra}}
\end{figure}

\begin{figure}
\begin{centering}
\includegraphics[width=1\columnwidth]{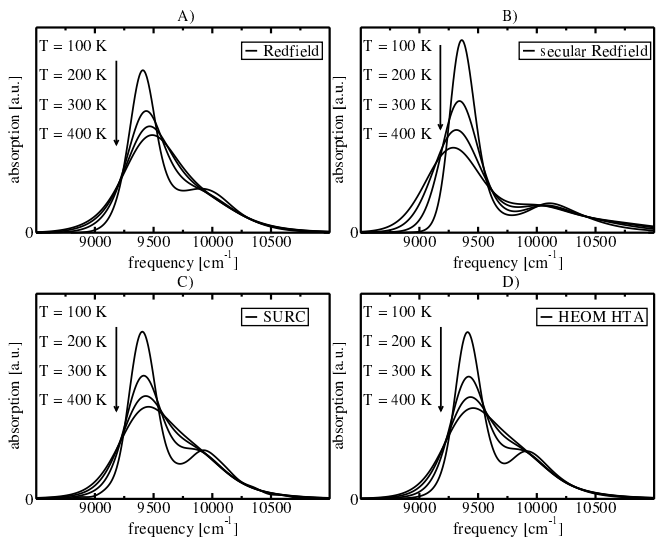}
\par\end{centering}

\caption{Temperature dependence of the absorption spectra of a molecular homodimer
coupled to an overdamped harmonic bath ($N=\infty$ in HOM) calculated
by A) Redfield theory, B) secular Redfield theory, C) SURC theory
and D) HEOM HTA. The parameters are $\epsilon_{1}=9600\;\mathrm{cm}^{\mbox{-}1}$,
$\epsilon_{2}=10000\;\mathrm{cm}^{\mbox{-}1}$, $\lambda_{1}=100\;\mathrm{cm}^{\mbox{-}1}$,
$\lambda_{2}=1000\;\mathrm{cm}^{\mbox{-}1}$, $J=300\;\mathrm{cm}^{\mbox{-}1}$
and $\tau_{C,1}=\tau_{C,2}=100\;\mathrm{fs}$. Absorption spectra
are calculated for temperatures ranging from 100 K to 400 K. \label{fig:Temperature-dependence}}
\end{figure}

\begin{figure}
\begin{centering}
\includegraphics[width=1\columnwidth]{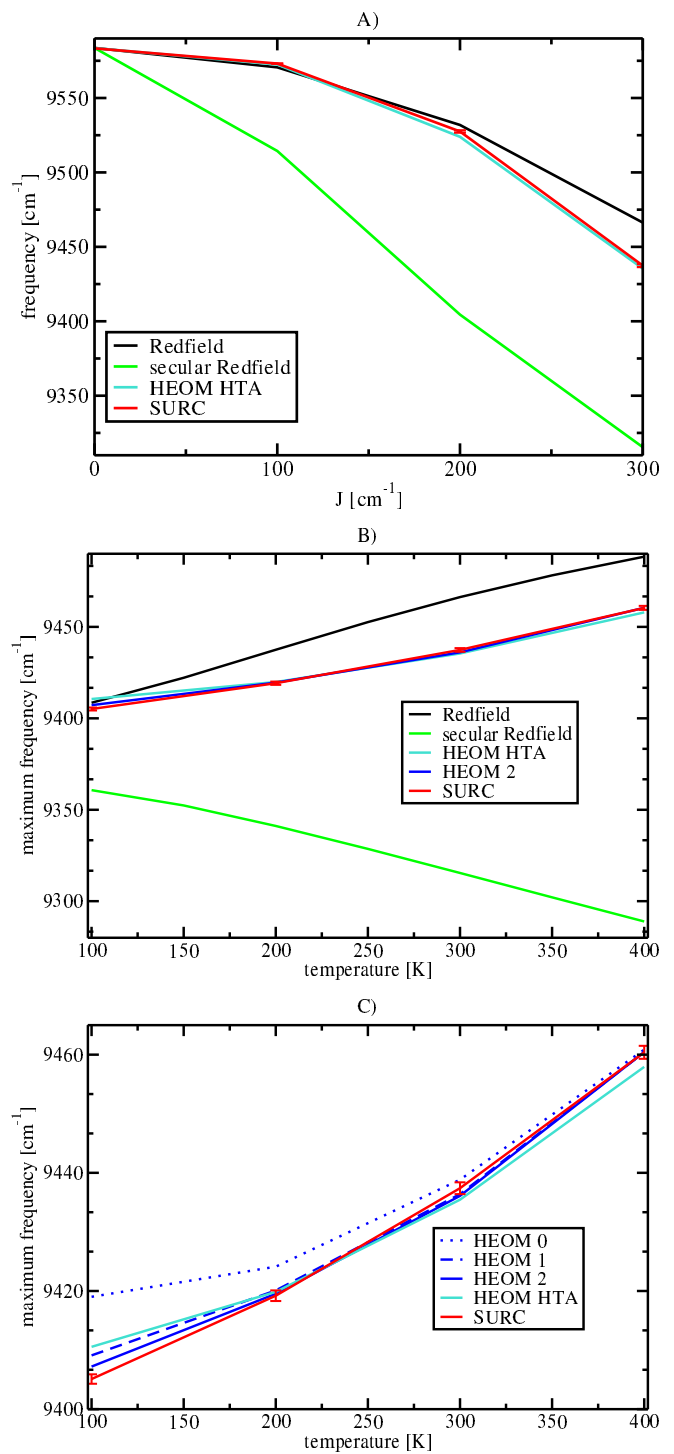}
\par\end{centering}

\caption{Positions of the maxima of the lower frequency peak of the spectra
for A) resonance coupling dependence (Fig.~\ref{fig:Absorption-spectra}),
B) temperature dependence (Fig.~\ref{fig:Temperature-dependence})
and detail of the temperature dependence showing comparison to various
approximations of HEOM. \label{fig:Peak-positions}}
\end{figure}

Figs.~\ref{fig:Absorption-spectra} and \ref{fig:Temperature-dependence}
present absorption spectra of the studied dimer system. The Fig.~\ref{fig:Absorption-spectra}
shows absorption spectra in three cases: $J=0\;\mathrm{cm}^{\mbox{-}1}$,
$100\;\mathrm{cm}^{\mbox{-}1}$ and $300\;\mathrm{cm}^{\mbox{-}1}$.
For $J=0\;\mathrm{cm}^{\mbox{-}1}$, this is a problem of non-interacting
monomers which is exactly solvable \cite{Doll2008a,Mancal2010c},
and all theories give the same result. With a gradual increase of
$J$, we see an increase of the excitonic splitting, noticeable as
a shift of the higher of the peaks (lower transition frequency) to
the red. The results of non-secular Redfield theory and SURC calculations
are similar, while the secular approximation gives an exaggerated
peak splitting. The HEOM, which is considered an exact theory in this
regime of parameters, is in very good agreement with the SURC method.
The Fig.~\ref{fig:Peak-positions}A plots the positions of the lower
frequency peak for the four theories and quantifies the difference
in its shift with $J$. In order to estimate the error of the SURC
method, we divided the total 10$^{10}$ trajectories of every calculation
into five independent simulation runs. For every run, we independently
calculated the absorption spectrum and extracted the peak position.
The error bars in Fig.~\ref{fig:Peak-positions} are calculated as
standard mean deviations of the peak position. Already here, we can
conclude that the non-secular Redfield theory captures the decrease
of the excitonic splitting with respect to the excitonic basis surprisingly
well. This can serve as a support for the conclusions of Refs.~\cite{Mancal2006b}
and \cite{Kuhn2002a} where non-secular effects in form of temperature
dependent band shifts and changes in fluorescence depolarization dynamics
were treated by Redfield theory.

\begin{figure}
\begin{centering}
\includegraphics[width=0.95\columnwidth]{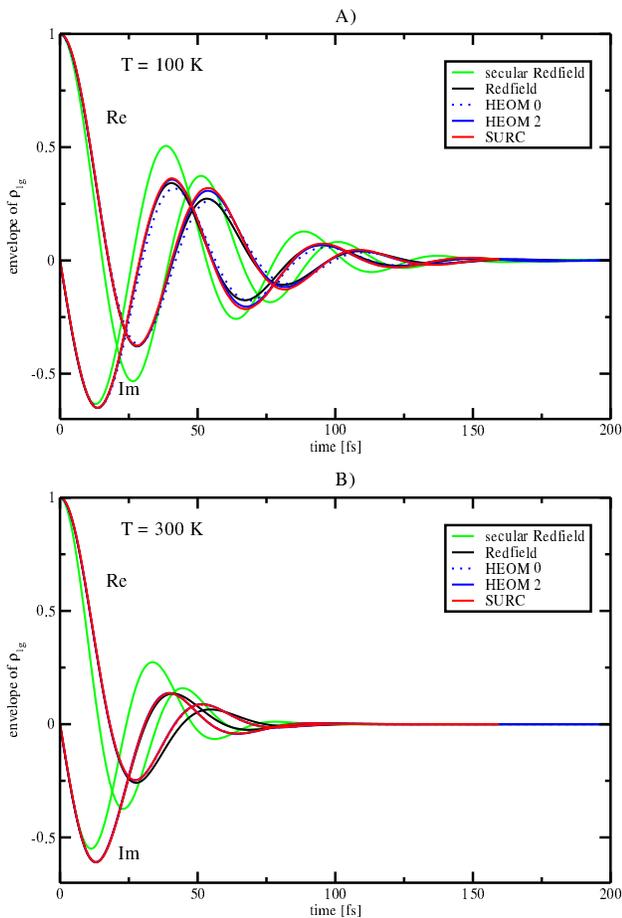}
\par\end{centering}

\caption{Dynamics of the optical coherence of the optically allowed site in
local basis for A) parameters from Fig.~\ref{fig:Temperature-dependence}A
($T=100\;\mathrm{K}$) and for B) parameters from Fig.~\ref{fig:Temperature-dependence}C
($T=300\;\mathrm{K}$). Frequency 10,000 $\mathrm{cm}^{\mbox{-}1}$
is subtracted in both calculations and the real (Re) and imaginary
(Im) parts of the evelope is plotted. In part~B, HEOM 0, HEOM 2 and
SURC overlap and HEOM dynamics is therefore not visible until the
end of SURC data in 160 fs. \label{fig:Dynamics}}
\end{figure}

In Fig.~\ref{fig:Temperature-dependence} we plot the temperature
dependence of the absorption spectra for the same model dimer. Here,
we can notice that the SURC is in nearly perfect match with the HEOM
HTA, and that the Redfield theory also captures the right tendency.
We can notice a shift of the lower frequency peak to higher frequencies
with increasing temperature. This demonstrates an increasing dynamic
localization by the bath with increasing temperature. The secular
Redfield gives opposite peak shift with respect to the other theories.
This shift is caused by the slight temperature dependent changes in
the lines shape, because the transition frequencies remain the same
for all temperatures. For lower temperatures, we can notice a difference
between the SURC and Redfield theory in the width of the wider peak.
With increasing temperature, these two theories also increasingly
differ in the lower frequency peak position, as is shown on Fig.~\ref{fig:Peak-positions}B.
As expected, the nearly perfect match of SURC and HEOM HTA gets worse
at lower temperatures. Calculation by HEOM with increasing number
of Matsubara frequencies shows a increasing degree of convergence
towards the SURC result. There is a small discrepancy of $2.1\;\mathrm{cm}^{-1}$
between the HEOM HTA and SURC in high temperatures, where the theories
should match. As the peak position is a very sensitive test, this
is probably caused by small numerical differences between the program
codes of HEOM $N_{\mathrm{Mat}}$ and HEOM HTA. As can be seen on
Fig.~\ref{fig:Dynamics}A, a difference on the same order between
HEOM 2 and SURC at $T=100\;\mathrm{K}$ visible on Fig.~\ref{fig:Peak-positions}C
is almost not recognizable in the dynamics of the envelope of the
optical coherence. 

In its present form, the SURC is limited to short time evolution due
to a breakdown of accuracy after the propagation time corresponding
roughly to 1.5 times the period of the oscillations caused by the
resonance coupling $J$. This is in turn caused by the dynamic sign
problem \cite{Makri1995} originating from the phase factor assigned
at each jump in a given trajectory. For the cases discussed here,
the GPU-HEOM implementation \cite{Kreisbeck2013Nanohub} exceeds the
SURC implementation in speed. However, the SURC is also massively
parallelizable. Moreover, it has only minimal memory requirements
independent of the system size and its computational cost does not
increase with the complexity of the EGCF. For example, the treatment
of spectral densities with $n$ shifted peaks corresponding to underdamped
vibrational modes by HEOM can be a challenging task in some cases
since the HEOM $N_{{\rm Mat}}$has computational difficulty growing
as $\frac{(N_{s}+N_{\mathrm{Mat}}N_{s}+k)!}{(N_{s}+N_{\mathrm{Mat}}N_{s})!k!}$
for EGCF of Eq.~(\ref{eq:EGCF-overdamped}) and $\frac{(2nN_{s}(1+N_{\mathrm{Mat}})+k)!}{(2nN_{s}(N_{\mathrm{Mat}}+1))!k!}$
for spectral density with $n$ shifted peaks. Computational costs
of HEOM HTA grow as $\frac{(2nN_{s}+k)!}{k!(2nN_{s})!}$ both in memory
and CPU time. Here $N_{s}$ is the number of sites and $k$ is the
Hierarchy depth required for convergence \cite{Tanimura1994,Kreisbeck2012a}.
In our calculations, typical hierarchy depth is $k=20$.  The SURC
method scales in CPU and memory proportionally to the dimension of
the reduced density matrix, which corresponds to $N_{s}$ for calculation
of the absorption spectra and to $N_{s}^{2}$ in the block of single
exciton states. Clearly, the initial cost of the SURC method is high
in current implementation. When a complex EGCF is required for the
description of a given bath, the cost of SURC calculation does not
increase. The SURC may therefore offer advantage over the HEOM for
larger systems with more involved spectral densities, or serve as
a good check of the HEOM convergence.

\section{Conclusion\label{sec:Conclusion}}

In this paper we proposed a Monte-Carlo method of evaluation of optical
coherence dynamics based on stochastic unraveling of resonance coupling
via cumulant expansion, which is exact for a harmonic bath. The method
can be easily used with a bath specified by an arbitrary energy gap
correlation function. We verified that the method correctly reproduces
the coherent dynamics, and that it gives correct dynamics for exactly
solvable model of molecular dimer with one harmonic oscillator coupled
to each molecule representing the bath. We calculated absorption spectra
of a model dimer for different values of a resonance coupling and
demonstrated that the proposed scheme gives results very similar to
the Hierarchical equations of motion. We investigated the temperature
dependence of the spectrum for the same dimer, finding again good
agreement between the Hierarchical equations of motion and the method
introduced in this paper. Comparison with non-secular Redfield theory
leads us to a conclusion that despite its well-known problems, such
as the positivity breakdown under some parameters, the Redfield theory
describes the non-secular effects in optical coherences rather well
for the typical parameters used in our work. 
\begin{acknowledgments}
This work was supported by the Czech Science Foundation (GACR), grant
no.~14-25752S. T.~K.~acknowledges funding by the DFG within the
Heisenberg program (KR 2889/5). C.~K.~acknowledges funding by the
Defense Advanced Research Projects Agency Award No.~N66001-10-4059.
Computational resources were provided by the MetaCentrum under the
program LM2010005 and the CERIT-SC under the program Centre CERIT
Scientific Cloud, part of the Operational Program Research and Development
for Innovations, Reg.~no.~CZ.1.05/3.2.00/08.0144.
\end{acknowledgments}
\bibliographystyle{prsty}
%\bibliography{main}

\end{document}